\documentclass[11pt,a4paper]{article}
\pdfoutput=1
 
\usepackage[margin=1in]{geometry}
 
\usepackage[T1]{fontenc}
\usepackage[utf8]{inputenc}
\usepackage{lmodern}
 
\usepackage{authblk}
\usepackage{graphicx}
\usepackage{amsmath,amssymb}
\usepackage{booktabs}
\usepackage{multirow}
\usepackage{placeins}
\usepackage{tikz}
\usetikzlibrary{arrows.meta,positioning,shapes.geometric}
\usepackage{capt-of}
\usepackage{url}
 
\usepackage[hidelinks]{hyperref}
\hypersetup{
  pdftitle={MagViT: Interpretable Multi-Magnification Transformers with Patient-Level Model Selection for Breast Histopathology},
  pdfauthor={Nabil Ashab, Soumit Kumar Kundu, Saif Mahmud Parvez, Shahadat Hossain Sohag, Bidhan Biswas, Nazmus Subha},
  pdfsubject={Breast histopathology; multi-magnification vision transformers},
  pdfkeywords={breast cancer, histopathology, vision transformer, multi-magnification, interpretability, BreakHis}
}

\title{MagViT: Interpretable Multi-Magnification Transformers with
Patient-Level Model Selection for Breast Histopathology}

\author[1]{Nabil Ashab\thanks{Accepted for publication in the International Conference on Electrical, Computer and Communication Technologies (ECCT 2026) proceedings by Taylor \& Francis Books. This is the author-produced version. The definitive Version of Record will be available via its DOI.}}
\author[1]{Soumit Kumar Kundu}
\author[1]{Saif Mahmud Parvez}
\author[1]{Shahadat Hossain Sohag}
\author[1]{Bidhan Biswas}
\author[2]{Nazmus Subha}
 
\affil[1]{Department of Computer Science and Engineering,
Dhaka International University, Dhaka, Bangladesh\protect\\
\footnotesize
\href{mailto:nabilashab@gmail.com}{nabilashab@gmail.com},
\href{mailto:soumitkumar4@gmail.com}{soumitkumar4@gmail.com},
\href{mailto:saif.mahmud.parvez@gmail.com}{saif.mahmud.parvez@gmail.com},\protect\\
\href{mailto:shahadat.h.sohag8@gmail.com}{shahadat.h.sohag8@gmail.com},
\href{mailto:bidhan.biswas970@gmail.com}{bidhan.biswas970@gmail.com}}
 
\affil[2]{Department of Microbiology, TMSS Medical College, Bogura, Bangladesh\protect\\
\footnotesize
\href{mailto:subhaimc7@gmail.com}{subhaimc7@gmail.com}}
 
\date{}
 
\begin{document}
\maketitle

\begin{abstract}
Breast cancer is one of the most common types of cancer among women around the world. Rapid detection and early treatment can hinder its progress to more complex stages and can impede its spread to other parts of the body. Histopathological image classification is the most common task in cancer detection due to its robustness in analyzing cellular data. Breast histopathology classification requires handling both multi-scale tissue morphology and clinically relevant generalization beyond the source domain. This paper presents MagViT, an interpretable multi-magnification transformer framework with scale-gated fusion and patient-level model selection. The model uses four BreakHis magnifications (40X, 100X, 200X, 400X) and extracts per-scale representations with a ViT backbone, and combines them via a learnable gate that masks missing scales. Patient-level five-fold cross-validation with a fixed seed has been run and compared with three architectural branches. The most accurate branch is then selected as the final model due to the strongest patient-level accuracy while retaining the simplest fusion pathway. On BreakHis, our architecture achieves a mean image accuracy of 0.9191, a mean patient accuracy of 0.9643, and a mean macro-F1 of 0.9042. External transfer experiments provide preliminary evidence of cross-dataset generalization under controlled adaptation settings on BUSI (image accuracy 0.8306, macro-F1 0.7480, patient accuracy 0.8291) and IDC (image accuracy 0.8577, macro-F1 0.8191, patient accuracy 0.8372). Grad-CAM visualization indicates that the model focuses on diagnostically significant and meaningful regions across magnifications. Relative to prior ViT-centered BreakHis work, this study emphasizes patient-level selection and cross-dataset robustness under a reproducible protocol.

\end{abstract}

\section{Introduction}
Breast cancer pathology is a high-impact diagnostic workflow where both sensitivity and specificity directly affect downstream treatment decisions \cite{acs2021,elston1991}. Histop\-athological image analysis has therefore become a major target for machine learning and deep learning research \cite{litjens2017,zhou2020}. In this setting, BreakHis is one of the most established datasets because it offers diverse patient samples and four magnification levels that expose multi-scale morphology \cite{spanhol2016}.

Transformer architectures, especially ViT, have recently become competitive in medical imaging tasks due to their ability to model long-range context \cite{dosovitskiy2020,xu2023}. Prior BreakHis-focused transformer studies, such as the ViT-DeiT ensemble approach \cite{alotaibi2023}, report strong performance and reinforce ViT viability in this domain. However, high image-level performance alone is not sufficient for practical reliability in clinical settings \cite{aggarwal2021}. Two gaps repeatedly appear in the literature and in reproduction practice.

Image-level aggregation often masks whether a model remains reliable at the patient level, where clinical decisions are actually made. Also, strong performance on BreakHis does not necessarily translate under distribution shifts, such as BUSI ultrasound patterns or invasive ductal carcinoma patches. In addition, interpretability is frequently limited to isolated examples, leaving consistency unclear. To address these issues, this study introduces MagViT, combining multi-scale fusion, patient-level selection, and comprehensive, artifact-driven evaluation.

\textbf{Main contributions.}
\begin{itemize}
\item A mask-aware multi-magnification transformer fusion framework over 40X, 100X, 200X, 400X views.
\item A patient-level model-selection protocol across three branch variants using five-fold CV.
\item Cross-dataset transfer evidence on BUSI and IDC with both linear-probe and few-shot adaptation.
\item Comprehensive visual diagnostics: training curves, ROC curves, confusion matrices, magnification importance, and Grad-CAM.
\end{itemize}

\section{Related Work}
Machine Learning and deep learning have been widely used to Histopathological image diagnosis for breast cancer, mainly due to the super complex tissue morphology and clinical importance. CNN-based models (e.g., ResNet, DenseNet, EfficientNet)  \cite{alhaija2020,sarvamangala2021} showed strong performance in classification, often combined with multi-scale fusion strategies, ex, Khan et al. \cite{khan2022}. Despite high accuracy, they are often computationally expensive and difficult to deploy in practice. Similarly, architectures like VGG networks integrated with attention mechanisms like CBAM have also been explored \cite{ijaz2023} to enhance feature representation. But their limitations remain the same, like high training complexity and limited efficiency, particularly when transfer learning is not fully utilized.

Multiple studies have leveraged pretrained CNNs for feature extraction followed by lightweight classifiers \cite{isewon2025,heikal2024}. For instance, combinations of DenseNet, NasNetMobile, and VGG16 have demonstrated strong performance on the BreakHis dataset. Attention mechanisms and models incorporating channel and spatial attention, such as CBAM and dual squeeze-and-excitation blocks, have further improved CNN-based pipelines \cite{cruzroa2014} by emphasizing diagnostically relevant regions. But many studies often only focus on a single magnification level (e.g., 40X or 400X), where multi-scale information is crucial for real-world diagnosis.

More recently, transformer-based architectures (ViTs), and a hybrid approach with CNN-transformer have also been explored. But still, many existing works emphasize image-level accuracy without considering patient-level evaluation, which is more clinically relevant. Also, interpretability and multi-domain robustness are often limited to qualitative examples, making it difficult to assess the consistency of model decisions. To address these limitations, our proposed framework combines transformer-based modeling with multi-magnification fusion, patient-level model selection, and transfer learning evaluation. Unlike prior works that prioritize peak accuracy, our approach emphasizes generalization, interpretability, and reproducibility, aligning more closely with real-world clinical requirements.

\section{Data and Experimental Protocol}
\makeatletter
\let\nobreak\relax
\makeatother
\subsection{Datasets}
In total, three datasets were used in this study. One source dataset for training and evaluate result, and the other two for the transfer learning study.

\textbf{BreakHis (source):} The primary dataset is Breast Cancer Histopathological Dataset - BreakHis \cite{spanhol2016}. It consists of  9,109 microscopic images from 82 patients of different magnification levels (40X, 100X, 200X, and 400X. It has 2,480 benign and 5,429 malignant samples. This is the main training dataset.

\textbf{BUSI (transfer):} Breast Ultrasound Images Dataset(BUSI) is a dataset created from 600 female patients between 25 and 75 years old \cite{aldhabyani2020}. with 780 images with an average image size of 500*500 pixels.

\textbf{IDC (transfer):} This dataset consists of diagnosed Invasive Ductal Carcinoma (IDCs) from 162 women at the Hospital of the University of Pennsylvania \cite{janowczyk2016}.

\subsection{Patient-Level Split Integrity}
Patient-level split integrity is very important because it mimics the actual clinical use. Often researcher ignore this fact and optimize their architecture for image-level split only. But the issue with this is image level split induce inflated result and mixes images from the same patient from training and test. It is actually one of the main causes for overfitting the training accuracy and not getting the intended result on the test.

The official fold metadata that came with the dataset was used to perform  5-fold cross-validation with a defined patient-level split. If leakage occurred, a check was enforced to abort execution and prevent patient overlap. Doing this, it was ensured that the same image doesn't appear in the training, validation, and test sets. This provides a more practical understanding of the model’s effectiveness in real-world clinical settings.

\section{MagViT Method}
\begin{figure}[ht]
    \centering
    \includegraphics[width=0.8\textwidth]{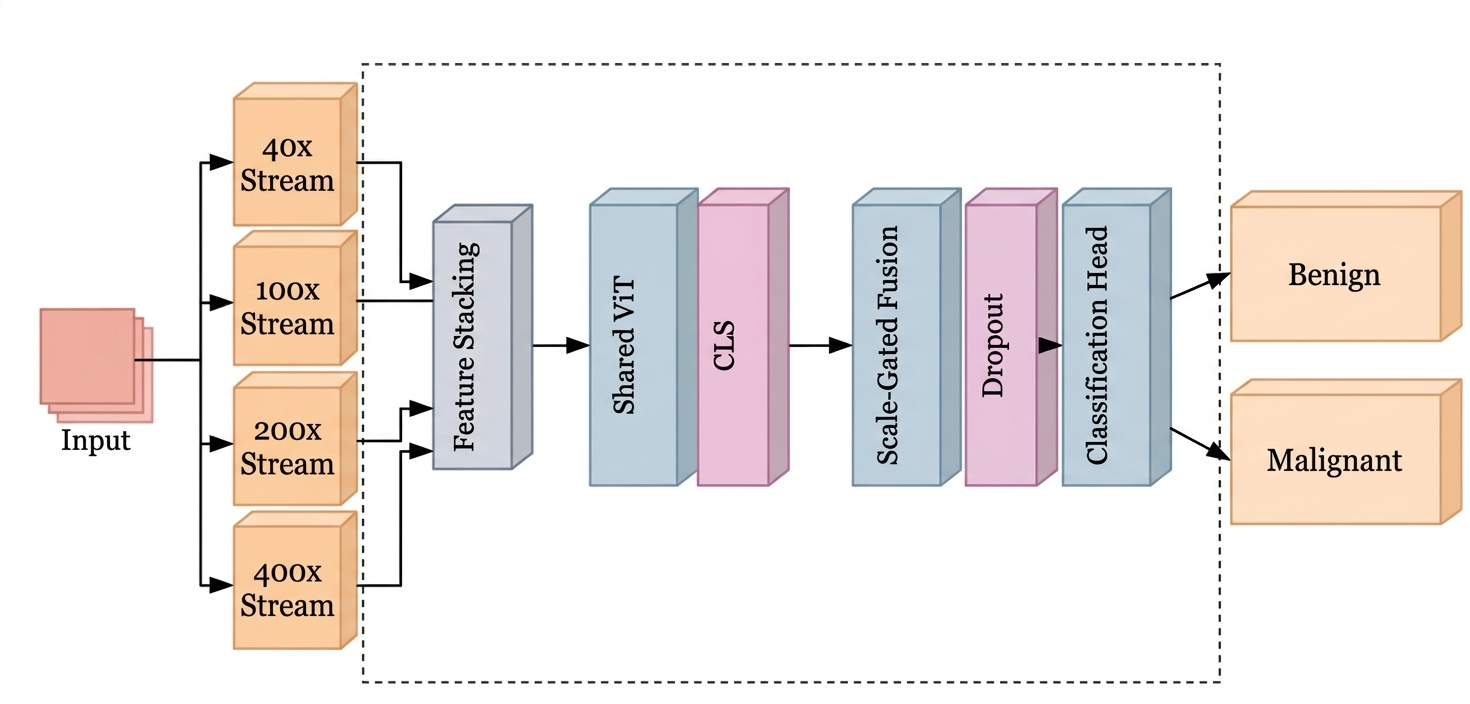}
    \caption{Proposed multi-magnification architecture with expanded streams and outputs}
    \label{fig:magvit_main}
\end{figure}

The proposed MagViT method is actually a novel pipeline selected from a series of pipelines after observing results and metrics from the whole experiment multiple times to address the real-world diagnosis and maximize the gain while doing it. The main part of the architecture is the same, multi-magnification feature extraction with a previously trained visual transformer base, fusing the extracted CLS tokens with scale-gated fusion, and then performing binary classification using a classifier head. This pathway is chosen after the experiment. Fig.\ref{fig:magvit_main} shows the full architecture of MagViT.

\subsection{Dataset Augmentation}
Augmentation includes random changes in the data, e.g., flipping the image horizontally or vertically (each with a 50\% chance), rotating it between -15° and +15°, adjusting brightness, contrast, saturation, and hue, and applying small shifts or size changes (up to 5\% in any direction and 5\% scale difference). Only one combination (randomly picked) is used for each group of training images. Exact same combination is then applied to every zoom level within the group. Although augmentation is used during the training phase, validation and test sets remain unaltered for clean and real output.

\subsection{Image Preprocessing}
A preprocessing pipeline was built to: i) preserve information and structural data of the same tissue regions across different magnification levels (40X, 100X, 200X, 400X), ii) keep the model resilient to rotation, color shifts, and positional changes. Sometimes, one or more zoom levels are missing. Then, missing zooms are marked using a mask. After loading the image, they are colored, resized, normalized, and converted to a tensor according to the ViT model’s requirements. Color jitter is also used for convenience.

\subsection{Reproducibility Preserved Core Settings}
These core settings are chosen to balance training stability and generalization. Excessive parameter tuning is avoided by these. Also, a fixed configuration across experiments ensures that performance differences arise from architectural design rather than hyperparameter optimization. All experiments were conducted using a fixed random seed of 42 to ensure consistent and reproducible results.

\begin{itemize}
    \item \textbf{Source task:} binary(benign vs malignant) BreakHis classification.
    \item \textbf{Split protocol:} patient-level 5-fold cross-validation using official \texttt{Folds.csv}.
    \item \textbf{Leakage control:} explicit train/test patient-overlap checks.
    \item \textbf{Backbone:} ViT-B/16 (\texttt{google/vit-base-patch16-224}) is used for global feature extraction \cite{dosovitskiy2020}.
    \item \textbf{Training defaults:} 15 epochs, learning rate $5 \times 10^{-5}$, cosine scheduler, warmup ratio 0.1, batch size 32, weight decay 0.01.
    \item \textbf{Loss:} class-weighted cross-entropy with label smoothing.
    \item \textbf{Metrics:} image accuracy, patient accuracy, macro-F1.
\end{itemize}

\subsection{Methodological and Fair-Comparison Rationale}

MagViT is designed to preserve patient-level reliability under realistic data constraints, reflecting the fact that clinical decisions operate at the patient rather than image level. Different magnification levels capture complementary morphological information. To capture the multi-scale nature of histopathology, a scale-gated fusion mechanism adaptively weights magnification-specific features, while masking prevents missing scales from introducing bias. Architectural comparison is kept controlled, with all branches trained under identical folds, seeds, and optimization settings. Hyperparameters remain fixed to isolate structural effects, enabling clearer interpretation of how design choices influence both image- and patient-level outcomes. Overall, the methodological design emphasizes controlled experimentation, robustness to missing data, and alignment with clinically relevant evaluation criteria.

\subsection{Multi-Magnification Representation}
Dataset consists of four magnifications $\mathcal{M}=\{40\mathrm{X},100\mathrm{X},200\mathrm{X},400\mathrm{X}\}$ so the magnification set $M=|\mathcal{M}|=4$, for patient region $i$ the multi-scale input is:
\begin{equation}
\mathbf{X}_i=\{\mathbf{x}_{i,m}\}_{m\in\mathcal{M}},
\label{eq:multi_input_set}
\end{equation}
where $\mathbf{x}_{i,m}\in\mathbb{R}^{H\times W\times 3}$ is the image at magnification $m$ when available.

An availability mask was introduced because some regions may miss one or more scales
\begin{equation}
s_{i,m}\in\{0,1\},\qquad s_{i,m}=1 \ \text{iff}\ \mathbf{x}_{i,m}\ \text{exists}.
\label{eq:availability_mask}
\end{equation}

Backbone is a single $f_{\text{ViT}}$ is implemented across all magnifications, which reduces branch-specific drift and also ensures a common feature geometry across multiple scales. For each available scale, the encoder output is summarized by the CLS token:
\begin{equation}
\mathbf{h}_{i,m}=f_{\text{ViT}}(\mathbf{x}_{i,m})_{\mathrm{CLS}}\in\mathbb{R}^{d}.
\label{eq:cls_token_feature}
\end{equation}
In this process, missing scales never contribute to the fused result, because unavailable scales are handled by the binary mask in fusion (Eq.~\ref{eq:availability_mask}).

\subsection{Scale-Gated Fusion}
In histopathology, cellular details can be seen at high magnification, and global tissue layout can be seen at low magnification \cite{khan2022}. Because of different magnifications, emphasizing different morphologies, an adaptive weight scale is needed to ensure a fixed averaging rule.

Concatenation of per-scale features
\begin{equation}
\mathbf{u}_i=[\mathbf{h}_{i,1};\mathbf{h}_{i,2};\cdots;\mathbf{h}_{i,M}] \in \mathbb{R}^{Md},
\label{eq:concat_features}
\end{equation}
and map them to gate logits
\begin{equation}
\mathbf{g}_i=\psi(\mathbf{u}_i),\qquad \mathbf{g}_i\in\mathbb{R}^{M},
\label{eq:gate_logits}
\end{equation}
where $\psi$ emphasizes a learnable gate network.

Masked softmax used for missing scale valid weighting:
\begin{equation}
\alpha_{i,m}
=
\frac{\exp(g_{i,m})\,s_{i,m}}
{\sum_{k=1}^{M}\exp(g_{i,k})\,s_{i,k}},
\qquad
\sum_{m=1}^{M}\alpha_{i,m}=1,\ \alpha_{i,m}\ge 0.
\label{eq:masked_softmax_gate}
\end{equation}
The fused descriptor is
\begin{equation}
\mathbf{z}_i=\sum_{m=1}^{M}\alpha_{i,m}\mathbf{h}_{i,m}.
\label{eq:fused_descriptor}
\end{equation}
Fusion is ensured both adaptive (sample-specific gates) and robust (strict exclusion of unavailable scales) by this implementation (Eqs.~\ref{eq:masked_softmax_gate} and \ref{eq:fused_descriptor}). This formulation can be interpreted as a constrained attention mechanism over magnification scales, where weights are dynamically assigned based on feature relevance. Unlike fixed averaging, this adaptive weighting improves robustness in heterogeneous samples and prevents dominance of noisy or less informative scales.

\subsection{Objective and Optimization}
Class-weighted cross-entropy is used with label smoothing. Let $y_i\in\{1,\dots,C\}$ be the class label ($C=2$). Now the smoothed target is
\begin{equation}
\tilde{y}_{i,c}=(1-\varepsilon)\,\mathbb{I}[c=y_i]+\frac{\varepsilon}{C}.
\label{eq:smoothed_target}
\end{equation}
The training objective is (With given predictive probabilities $p_{i,c}$ and class weights $w_c$)
\begin{equation}
\mathcal{L}
=
-\frac{1}{N}\sum_{i=1}^{N}\sum_{c=1}^{C}
w_c\,\tilde{y}_{i,c}\log p_{i,c}.
\label{eq:weighted_ce_loss}
\end{equation}
AdamW with cosine learning-rate decay as well as warmup is used for Optimization. Class-imbalance aware and stable transformer fine-tuning while keeping the objective compact is ensured with this setup (Eqs.~\ref{eq:smoothed_target} and \ref{eq:weighted_ce_loss}).

\subsection{Branch Definitions and Best Branch Selection}
\begin{figure}[ht]
    \centering
    \includegraphics[width=0.8\textwidth]{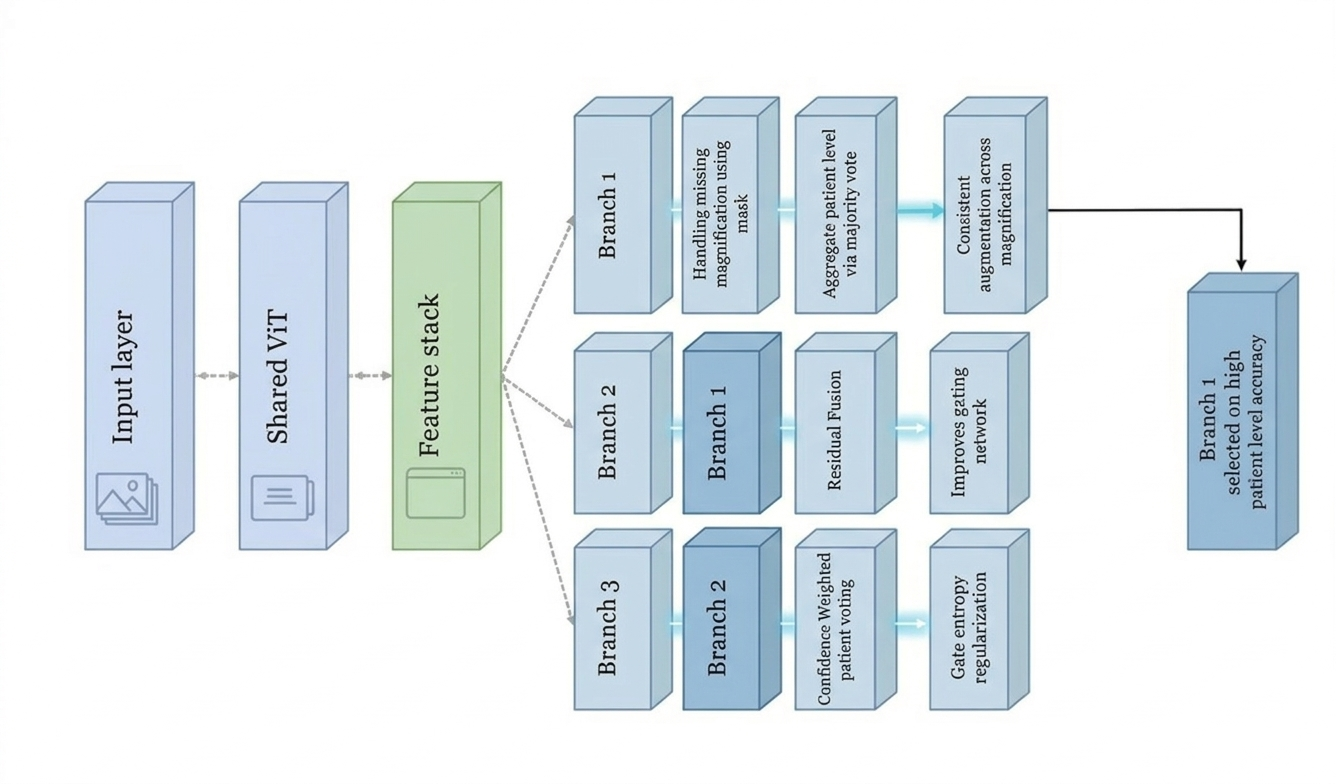}
    \caption{Branch-wise segregation and branch 1 selection for the main MagViT architecture}
    \label{fig:magvit_branch}
\end{figure}

Different branches are introduced to isolate the effects of fusion complexity and optimization policy while keeping hyperparameters and data splits unchanged. This setup enables controlled analysis of architectural variations without compromising classification performance, with core settings preserved across all branches. Inspection of training metrics and checkpoints reveals frequent patient-level decision flipping, motivating targeted adjustments categorized as critical, high, and optional. These modifications are organized into branches to simplify evaluation and track their impact over time, as illustrated in Fig.~\ref{fig:magvit_branch}.

\textbf{Branch 1} introduces mask-based missing-scale handling, consistent augmentation over different magnification levels, and majority-vote patient aggregation. 

\textbf{Branch 2} extends Branch 1 with residual fusion, normalization, and enhanced gating. 

\textbf{Branch 3} further adds confidence-weighted voting, partial backbone freezing, and entropy regularization.

Branch selection is based on a predefined patient-centric evaluation criterion that reflects real-world diagnosis scenarios. Even if Branch 2 shows marginal gains in image-level metrics, Branch 1 Table~\ref{tab:branch-main} demonstrates more consistent patient-level performance and lower architectural complexity. Branch 3 brings maximum architectural complexity with the lowest gain among others. Since minimizing patient-level misclassification rather than optimizing isolated predictions and balancing between stability, interpretability, and computational efficiency are the criteria that's why, Branch 1 is selected as the final configuration.

\subsection{Algorithm Description}
The algorithm of the full architecture is summarized below:
\begin{table}[ht]
\centering
\caption{Full End-to-End Final Pipeline}
\label{alg:magvit_pipeline}
\begin{tabular}{l}
\hline
\textbf{Step 1:} Create multi-magnification samples as groups from dataset \\
\textbf{Step 2:} Ensure patient-wise disjoint splits across folds \\
\textbf{Step 3:} Initialize MagViT model with shared encoder \\
\hline
\textbf{For each fold:} \\
\quad Apply identical augmentation across all magnifications \\
\quad Extract features from available scales \\
\quad Handle missing scales using availability mask \\
\quad Fuse features via simple aggregation pathway \\
\quad Train model under a consistent protocol \\
\quad Save fold-level metrics and diagnostics \\
\hline
\textbf{Step 4:} Aggregate predictions using majority voting \\
\textbf{Step 5:} Validate using patient-level evaluation \\
\textbf{Step 6:} Evaluate final model on test set \\
\hline
\end{tabular}
\end{table}

\section{Results on BreakHis}
\begin{figure}[ht]
\centering
\includegraphics[width=1.0\textwidth]{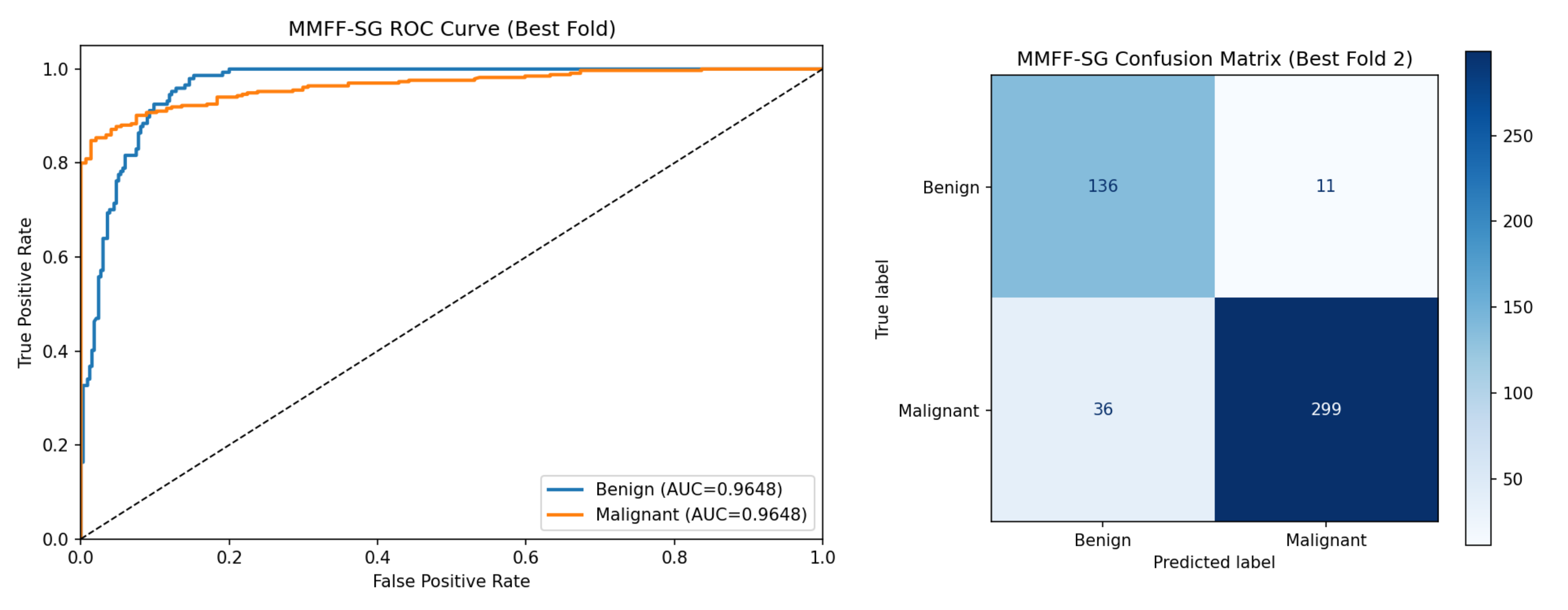}
\caption{In-domain binary diagnostics: ROC curve (left) and confusion matrix (right).}
\label{fig:in-domain-diag}
\end{figure}

The ROC curves and confusion matrices (Fig.\ref{fig:in-domain-diag}) indicate strong class separability and balanced performance across benign and malignant cases.

\subsection{Primary Branch Comparison and Ablation Study}
Table~\ref{tab:branch-main} summarizes branch-wise outcomes under a controlled setup. Evidence shows that Branch 1 achieves the highest patient-level mean accuracy (0.9643) and also aligns with the evaluation priority. Branch 2 has slightly higher image-level accuracy but patient-level performance declines here. Branch 3, on the other hand, has maximum architectural complexity with overall accuracy reduced results. Based on these observations, Branch 1 is selected for integration into the main architecture.

Complementary analysis in Table~\ref{tab:ablation-main} highlights the role of training components. Removing augmentation, color jitter, or class weighting consistently lowers patient-level accuracy. This shows their contribution to robustness and generalization. The current ablation goes towards isolated branch-level architectural effects. More fine-grained component comparisons—such as alternative fusion strategies—remain limited in this version and are deferred as future work. That's why Results are interpreted as controlled, relative comparisons rather than absolute superiority.

\begin{table}
\centering
\caption{Component-level ablation study (patient-level protocol retained).}
\label{tab:ablation-main}
\begin{tabular}{lcccc}
\toprule
Setting & Img. Acc. Mean & Img. Acc. Std & Pat. Acc. Mean & Pat. Acc. Std \\
\midrule
No augmentation & 0.9015 & 0.0381 & 0.9357 & 0.0539 \\
No color jitter & 0.9103 & 0.0359 & 0.9589 & 0.0467 \\
No class weights & 0.9094 & 0.0354 & 0.9482 & 0.0511 \\
\bottomrule
\end{tabular}
\end{table}

\subsection{Contextual Comparison with Prior Work}
Table~\ref{tab:patient-benchmark} compares the proposed approach with representative methods on the BreakHis dataset. The existing results in prior CNN-based and hybrid models in general fall within a comparable range. Most pipelines rely on single-scale inputs or image-level aggregation. Also, Differences in data splits, preprocessing, and evaluation protocols introduce variance. That's why a cautious interpretation of direct numerical comparisons is required here.

MagViT, in contrast, emphasizes patient-level evaluation through multi-magnification fusion and achieves a patient-level accuracy of 0.9643. This output indicates competitive performance and offers a closer alignment with clinically relevant assessment criteria, even if strict equivalence across studies cannot be assumed.

\begin{table}
\centering
\caption{Patient-level accuracy comparison on BreakHis (reported results from prior studies; not strictly comparable due to protocol differences).}
\label{tab:patient-benchmark}
\begin{tabular}{lcc}
\toprule
\textbf{Method} & \textbf{Patient Acc.} & \textbf{Notes} \\
\midrule
ResNet-18 (single-scale) & 0.8571 & Standard CNN baseline \\
ResNet-50 (single-scale) & 0.8929 & Deeper CNN variant \\
CNN multi-scale fusion & 0.9286 & Feature-level aggregation \\
AutoML (neural search) & 0.9143 & Architecture search baseline \\
Hybrid CNN ensemble & 0.9357 & Multi-backbone ensemble \\
\midrule
\textbf{MagViT (proposed)} & \textbf{0.9643} & MMFF-ViT \\
\bottomrule
\end{tabular}
\end{table}

\begin{table}
\centering
\caption{Branch comparison on BreakHis (patient-level 5-fold CV).}
\label{tab:branch-main}

\setlength{\tabcolsep}{10pt}

\begin{tabular}{lccc}
\toprule
Branch & Image Acc. & Patient Acc. & Macro-F1 \\
\midrule
Branch 1 & 0.9191 $\pm$ 0.0335 & 0.9643 $\pm$ 0.0357 & 0.9042 $\pm$ 0.0379 \\
Branch 2 & 0.9238 $\pm$ 0.0378 & 0.9571 $\pm$ 0.0466 & 0.9090 $\pm$ 0.0482 \\
Branch 3 & 0.8260 $\pm$ 0.0232 & 0.9000 $\pm$ 0.0299 & 0.8005 $\pm$ 0.0288 \\
\bottomrule
\end{tabular}
\end{table}

\subsection{Statistical and Robustness Interpretation}
Branch metrics are reported as mean $\pm$ standard deviation across five folds. For Branch 1, approximate 95\% confidence intervals are: image accuracy $0.9191 \pm 0.0416$, patient accuracy $0.9643 \pm 0.0443$, and macro-F1 $0.9042 \pm 0.0470$, offering a clearer view of variability beyond point estimates. At the fold level, patient accuracy shows four ties and one advantage for Branch 1 over Branch 2, suggesting limited separability under $n=5$. Image accuracy and macro-F1 favor Branch 2 in three folds. A paired Wilcoxon signed-rank test yields $p=0.8125$ (image accuracy), $p=1.0000$ (patient accuracy), and $p=0.8125$ (macro-F1), indicating no significant difference. Observed ranges—$0.8813$ to $0.9659$ (image) and $0.8571$ to $1.0000$ (patient)—reflect fold variability and justify reporting both central tendency and dispersion, which reflects robustness also. Given the small sample size, statistical power remains constrained; model selection, therefore, follows the predefined patient-centric criterion.

\subsection{Fold-Level Stability and Magnification-Wise Analysis}
This is an important part of the evaluation because it shows not only fold-wise patient-level accuracy but also shows that the performance remains stable across the official five folds. Table~\ref{tab:fold-wise-full} reports fold-wise metrics for the selected configuration. Here, 2 and 4 fold suffers the most because of the missing magnifications for some patients, while 3 and 5 thrived because of the balanced dataset on these folds.

Table~\ref{tab:magwise} summarizes single-magnification fold-level behavior, where it shows 200X and 100X are consistently strong; 400X exhibits more variance. This result actually shows motivation towards multi-scale fusion.

\begin{table}
\centering
\caption{Fold-wise performance of selected full model (Branch-1-compatible final pipeline).}
\label{tab:fold-wise-full}

\setlength{\tabcolsep}{9pt}
\renewcommand{\arraystretch}{1.15}

\begin{tabular}{lcccc}
\toprule
Fold & Image Acc. & Patient Acc. & Macro-F1 & Runtime (min) \\
\midrule
1 & 0.9243 & 0.9286 & 0.9112 & 39.36 \\
2 & 0.8959 & 0.8571 & 0.8728 & 40.89 \\
3 & 0.9659 & 1.0000 & 0.9585 & 40.15 \\
4 & 0.8813 & 0.8571 & 0.8645 & 40.26 \\
5 & 0.9629 & 1.0000 & 0.9569 & 38.80 \\
\bottomrule
\end{tabular}
\end{table}

\begin{table}
\centering
\caption{Magnification-wise summary (mean across 5 folds).}
\label{tab:magwise}
\begin{tabular}{lccc}
\toprule
Magnification & Mean Image Acc. & Mean Patient Acc. & Mean Macro-F1 \\
\midrule
40X & 0.9106 & 0.9786 & 0.8972 \\
100X & 0.9112 & 0.9643 & 0.8998 \\
200X & 0.9209 & 0.9643 & 0.9067 \\
400X & 0.8920 & 0.9357 & 0.8795 \\
\bottomrule
\end{tabular}
\end{table}
\subsection{Transfer Learning Results}
This section examines cross-dataset generalization under controlled, lightweight adaptation rather than positioning the study as a full domain adaptation benchmark. The evaluation focuses on how learned representations transfer across imaging modalities using linear probing and few-shot learning. Experiments on BUSI dataset and IDC dataset are summarized in Table~\ref{tab:transfer-main}, where few-shot learning consistently improves patient-level performance.

\begin{table} [ht]
\centering
\caption{Transfer performance on BUSI and IDC.}
\label{tab:transfer-main}
\begin{tabular}{llcccc}
\toprule
Dataset & Setting & Image Acc. & Macro-F1 & Patient Acc. & ROC-AUC \\
\midrule
BUSI & Linear probe & 0.7581 & 0.4312 & 0.7436 & 0.6908 \\
BUSI & Few-shot & 0.8306 & 0.7480 & 0.8291 & 0.8610 \\
IDC & Linear probe & 0.8503 & 0.8194 & 0.7907 & 0.9144 \\
IDC & Few-shot & 0.8577 & 0.8191 & 0.8372 & 0.9197 \\
\bottomrule
\end{tabular}
\end{table}

The analysis excludes specialized domain adaptation techniques such as adversarial alignment, discrepancy minimization, and test-time adaptation, and does not incorporate multi-center external cohorts. As a result, these findings are best interpreted as preliminary evidence of cross-dataset generalization rather than a definitive comparison with state-of-the-art domain adaptation approaches.

\begin{figure}
\centering
\includegraphics[width=1.0\linewidth]{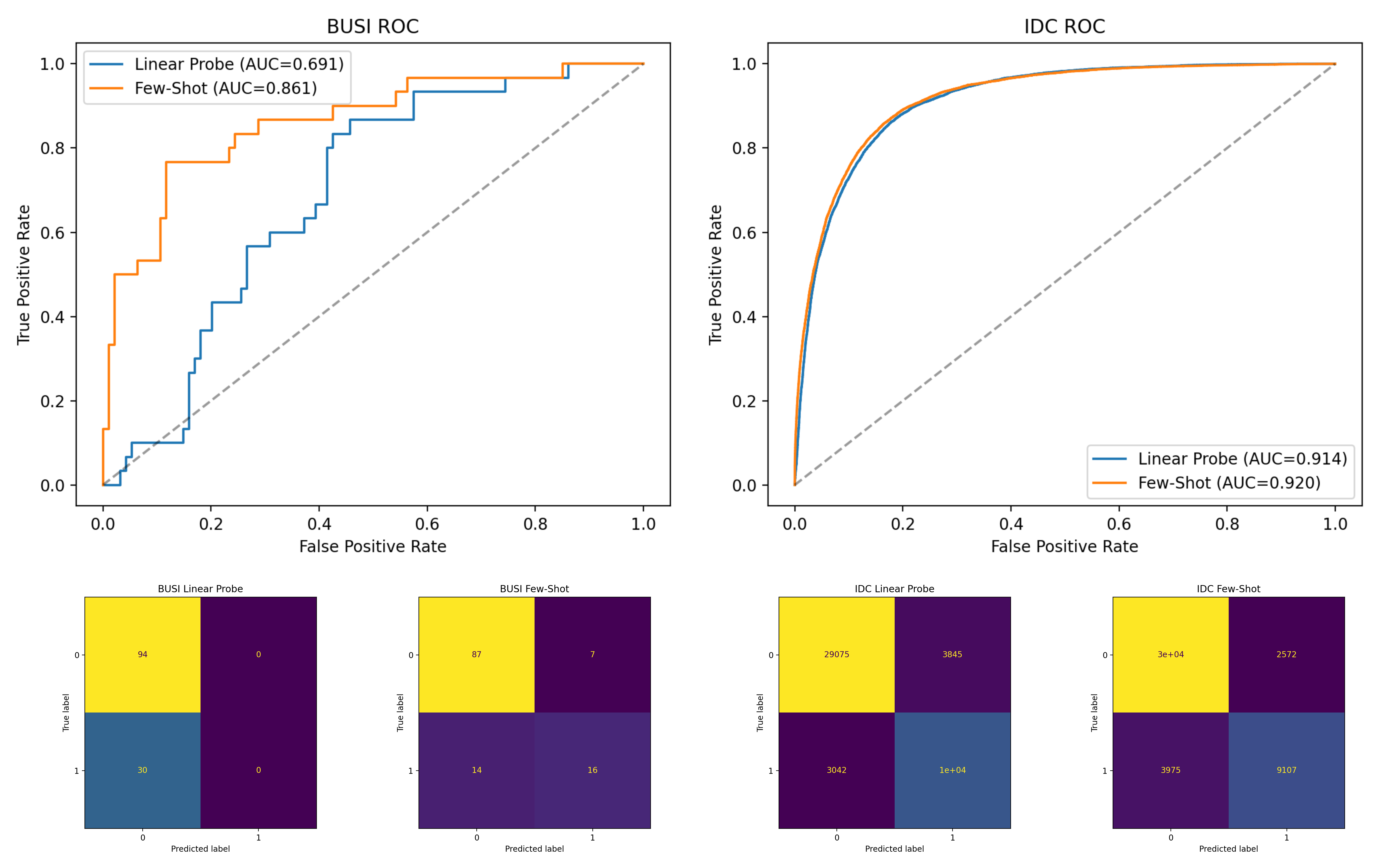}
\caption{BUSI (left) and IDC (right) transfer diagnostics: ROC curve and confusion matrix.}
\label{fig:busi-idc}
\end{figure}

Fig.~\ref{fig:busi-idc} illustrates model performance on the BUSI dataset and IDC dataset. On BUSI, few-shot learning achieves a higher AUC (0.861 vs 0.691), indicating improved class separation, which is also reflected in confusion matrices with fewer misclassifications than the linear probe. For IDC, AUC values remain close, showing only a slight advantage for few-shot learning, while confusion matrices indicate consistently high prediction accuracy with minimal errors.

\section{Grad-CAM Visualization}
\begin{figure}[ht]
\centering
\includegraphics[width=0.9\textwidth]{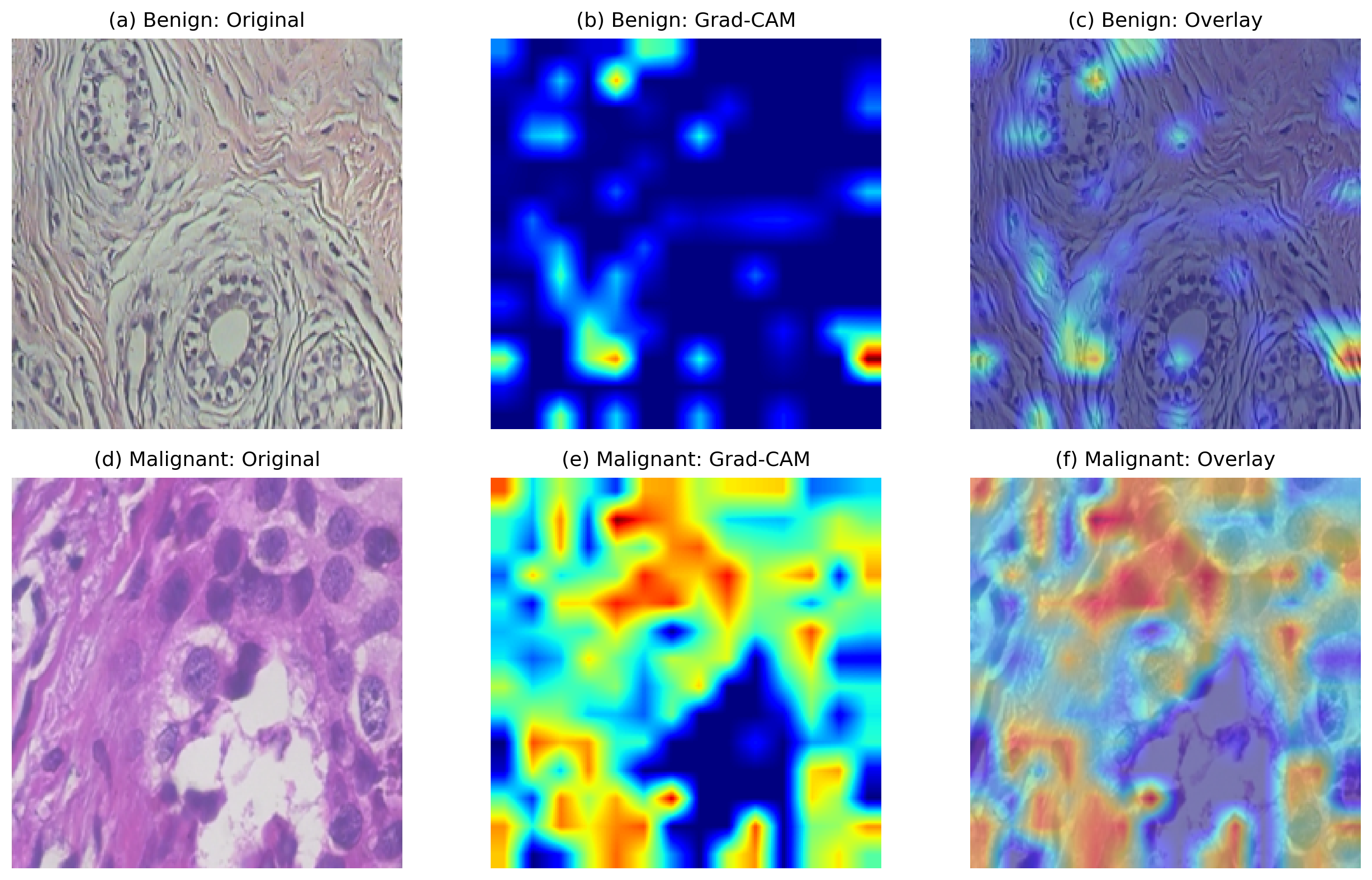}
\caption{Grad-CAM visualization for benign and malignant samples. Warmer colors indicate a higher contribution to the predicted class.}
\label{fig:gradcam-main}
\end{figure}

For explainability, Grad-CAM is used to highlight image regions that match the prediction. In here it is computed on a convolution layer (ViT projection layer where patch embedding happens). Upsampling and overlaid is then used on the histopathology image from BreakHis. Two correct predictions are shown in Fig.\ref{fig:gradcam-main}. One for benign and one for malignant. Original image, Grad-CAM heatmap, and heatmap overlay are also depicted for both predictions. Here, high-response regions indicate diagnostically influential tissue patterns used by the model.

\section{Discussion}
The branch comparison indicates a non-trivial trade-off between raw image-level optimization and patient-level stability. Branch 2 slightly outperforms Branch 1 on image accuracy and macro-F1. On the other hand, simply designed Branch 1 remains better on patient accuracy. This work enhances patient-level selection and adding transfer-centered stress tests. For real-world deployments where patient-level consistency is prioritized,  Branch 1 is more favorable. This focus leads to more conservative but operationally informative conclusions: strong in-domain performance is necessary, but robust cross-dataset behavior and transparent diagnostics are essential for broader reliability. Three practical observations emerge from this:
\begin{itemize}
    \item Multi-scale fusion improves resilience even when one magnification carries weak signal.
    
    \item Patient-level aggregation reduces overconfidence in isolated image-level mistakes.
    
    \item Few-shot adaptation is especially beneficial for BUSI, where modality shift (from histology to ultrasound) is substantial.
\end{itemize}
The transfer analysis reinforces that adaptation strategy matters. On BUSI, few-shot improves both discrimination and class balance quality (macro-F1) as the ultrasound image dataset is a bit different than Histopathological ones. On IDC, gains are smaller but still positive for patient accuracy and ROC-AUC, suggesting the source representation is already strong and fine-tuning acts as calibration. Although the results show consistent patterns across folds, the small number of folds and datasets limits the statistical strength of the conclusions. Broader validation with larger, multi-center cohorts and repeated cross-validation would provide a more reliable assessment of robustness.

\section{Limitations and future directions}
Source training remains binary and does not cover subtype-level pathology labels. External transfer is limited to BUSI and IDC; broader multi-center testing is still required. Calibration, uncertainty decomposition, and prospective study design are not yet included. As further work,  confidence calibration and threshold analysis can be added at patient-level. This can be extended to weakly supervised whole-slide settings. Domain adaptation strategies under staining and scanner variation can be added. The current evaluation is restricted by the scale of available data and the number of validation folds, which may affect the generalizability of the findings. Expanding the study to include diverse, multi-center datasets and more extensive validation protocols would help strengthen confidence in the results.

\section{Conclusion}
MagViT, the proposed architecture in this paper  combines different zoom-level ViT representations with mask-aware scale-gated fusion and patient-level branch selection. The selected Branch 1 configuration delivers strong BreakHis patient-level performance. This architecture also shows generalization to ultrasound image dataset (BUSI) and another Histopathological dataset (IDC). The detailed tables and diagnostic figures, supports a practical conclusion: patient-level model selection and transfer-aware evaluation provide a more reliable path than image-level optimization alone. This study highlights the importance of patient-level evaluation and controlled multi-scale fusion in improving the practical reliability of Histopathological classification systems.

\end{document}